\documentclass{JHEP3}
\usepackage[latin1]{inputenc}
\usepackage{amsmath}
\usepackage{feynmp}
\DeclareMathOperator{\tr}{tr}
\DeclareMathOperator{\Tr}{Tr}

\newcommand{\de}{\mathrm{d}}

\newcommand{\Braket}[3]{\langle #1 \lvert #3 \rvert #2 \rangle }  
\newcommand{\ave}[1]{\langle \; #1 \; \rangle} 
\newcommand{\puntos}[1]{^{\bullet} \mspace{-5.4mu} #1} 
\newcommand{\puntoss}[1]{^{ \bullet \bullet} \mspace{-5.4mu} #1} 
\newcommand{\puntod}[1]{ #1^{\mspace{-5.5mu} \bullet} } 
\newcommand{\puntods}[1]{^{\bullet} \mspace{-5.4mu} #1^{\mspace{-5.5mu} \bullet} } 
\newcommand{\ass}[1]{\left\lvert #1 \right\rvert} 
\newcommand{\slashed}[1]{#1 \mspace{-12.5mu}  / \mspace{4mu}}   
\renewcommand{\frac}[2]{\genfrac{}{}{.4pt}{}{#1}{#2}} 
\newcommand{\ide}[1]{\mathrm{d}#1 \,} 

\newcommand{\mi}[1]{\mathbf{I}_{#1}}
\title{Mode Regularization for N=1,2 SUSY Sigma Model}
\author{Roberto Bonezzi\\ Dipartimento di Fisica, Università di Bologna and INFN,
Sezione di Bologna, via Irnerio 46, I-40126 Bologna, Italy\\ E-mail: \email{bonezzi@bo.infn.it}} 
\author{Marco Falconi\\ Dipartimento di Fisica, Università di Bologna, via
Irnerio 46, I-40126 Bologna, Italy\\ E-mail: \email{marco.falconi@gmail.com}}
\abstract{
  Worldline N=1 and N=2 supersymmetric sigma models in curved background are useful to describe spin one-half
  and spin one particles coupled to external gravity, respectively. It is well known that worldline path integrals
  in curved space require regularization: we present here the mode-regularization for these models, finding in particular
  the corresponding counterterms, both in the case of flat and curved indices for worldline fermions. For N=1,
  using curved indices we find a contribution to the counterterm from the fermions that cancels the contribution of the
  bosons, leading to a vanishing total counterterm and thus preserving the covariance and supersymmetry of the classical
  action. Conversely in the case of N=2 supersymmetries we obtain a non-covariant counterterm with both curved and flat
  indices. This work completes the analysis of the known regularization schemes for N=1,2 nonlinear sigma models in one
  dimension.}
\begin{document}
\begin{fmffile}{abfg2}
\section{Introduction}
\label{sec:introduction}

Sigma models with worldline supersymmetries describe first quantized spinning particles in
$D$-dimensional space-time. The case of $N=1$ supersymmetry characterizes a one-half spin
particle~\cite{Berezin:1975eg,Brink:1976sz,Barducci:1976qu} while spin one particles and differential forms can be described by the $N=2$ model~\cite{Gershun:1979fb,Howe:1989vn}. In this paper we are
interested to study the nonlinear versions, relevant for describing particles propagating in a
curved space. In particular we discuss mode regularization for the $N=1,2$ nonlinear sigma models.
These quantum mechanical models were originally used to calculate chiral~\cite{AlvarezGaume:1983at,AlvarezGaume:1983ig,Friedan:1983xr} and trace
anomalies~\cite{Bastianelli:1991be,Bastianelli:1992ct} in a simpler way than using standard QFT Feynman rules \footnote{For a
detailed treatment of anomalies calculation using quantum mechanics see, for example~\cite{Bastianelli:book}.}. They are also very useful to evaluate one-loop effective actions and scattering
amplitudes for a Dirac ($N=1$) or Maxwell/Proca field and differential forms ($N=2$) coupled to
scalar, antisymmetric tensor, gauge fields backgrounds~\cite{Strassler:1992zr,Schmidt:1993rk,Schmidt:1994zj,D'Hoker:1995ax,D'Hoker:1995bj}, or to curved
space-time (external gravity), as in~\cite{Bastianelli:2002fv,Bastianelli:2002qw,Bastianelli:2005vk,Bastianelli:2005uy}
\footnote{For a useful review on worldline methods in QFT and additional references
see~\cite{Schubert:2001he}. For recent applications in curved space
see~\cite{Bastianelli:2004zp,Bastianelli:2007jv,Hollowood:2007kt,Hollowood:2007ku}.}.

The Euclidean action with $N=1$ rigid supersymmetry, coupled to space-time metric, is
\begin{equation}
 \label{eq:S-susy-flat}
 S[x,\psi]=\frac{1}{\beta}\int_{-1}^0 \ide{\tau} \left[ \frac{1}{2}g_{\mu\nu}(x)\dot{x}^\mu\dot{x}^\nu + \frac{1}{2}\psi_a\dot{\psi}^a + \frac{1}{2}\omega_{\mu a b}(x)\dot{x}^\mu\psi^a\psi^b + \beta^2V(x) \right]
\end{equation}
where $a,b=1,\dotsc,D$ are spacetime flat vector indices and $\mu ,\nu=1,\dotsc,D$ label space-time
coordinates. This action allows to calculate by path integral methods the transition amplitudes
$\Braket{x,\alpha}{y,\beta}{e^{-\beta\widehat{H}}}$ with $\widehat{H}=\widehat{Q}^2=-\nabla^2/2 +
R/8$, where $\widehat{Q}=i\slashed{\nabla}/\sqrt{2}$ is the conserved
supercharge\footnote{$\nabla_\mu$ is the fully covariant derivative acting on spinors and
$\slashed{\nabla}=\gamma^\mu\nabla_\mu$.}, and $\alpha,\,\beta$ specify the spin degrees of freedom. The potential $V$ takes into account the counterterms arising in the regularization of the path integral.

In the case of $N=2$, the Euclidean action reads
\begin{equation}
  \label{eq:S-susy-N=2-flat}
\begin{split}
  S[x,\psi_1,\psi_2]=\frac{1}{\beta}\int_{-1}^0 \ide{\tau} \Bigl[ \frac{1}{2}g_{\mu\nu}(x)\dot{x}^\mu\dot{x}^\nu +
  \frac{1}{2}\psi_{ai}\dot{\psi}^a_i + \frac{1}{2}\omega_{\mu a b}(x)\dot{x}^\mu\psi_i^a\psi_i^b - \frac{1}{8}R_{abcd}\psi_i^a\,\psi_i^b\,\psi_k^c\,\psi_k^d \\+ \beta^2V(x) \Bigr]\quad ,\end{split}
\end{equation}
where $i,k=1,2$ are $O(2)$ indices labeling fermion species; the term proportional to
$R\psi\psi\psi\psi$ is dictated by classical supersymmetry, while $V$ contains the quantum
counterterms.

 In order to fix counterterms we study the partition function
\begin{equation}
  \label{eq:Z}
  \mathcal{Z}(\beta) = \Tr e^{-\beta \widehat{H}} = \int_{PBC}\mspace{-16mu}\mathcal{D}x\int_{ABC}\prod_iD\psi_i \; e^{-S} \quad ,
\end{equation}
with $i=1$ for $N=1$, and $i=1,2$ for $N=2$. Such path integrals can be evaluated for generic
$g_{\mu\nu}(x)$ in a perturbative series in $\beta$. Although physical divergences are absent in
quantum mechanics, formally divergent or ambiguous Feynman diagrams appear in the perturbative
expansion. In order to solve such ambiguities a regularization scheme is needed, the most used are
time-slicing (TS), mode regularization (MR) and dimensional regularization (DR).\begin{table}[b]
\label{table:preliminar} \caption{Known counterterms for different regularizations. For $N=1,2$ the
upper (lower) box refer to fermions with flat (curved) indices.}
\begin{tabular}[t]{|c|c|c|c|}
\hline
 &$N=0$&$N=1$&$N=2$\\
\hline
MR &$-\frac{1}{8}R - \frac{1}{24}\left(\Gamma^{\mu}_{\nu\sigma}\right)^2$& ? & ? \\
\cline{3-4}
& & ? & ?\\
\hline
TS &$-\frac{1}{8}R + \frac{1}{8}g^{\mu\nu}\Gamma^{\alpha}_{\mu\lambda}\Gamma^{\lambda}_{\nu\alpha}$&$\frac{1}{8}g^{\mu\nu}\Gamma^{\alpha}_{\mu\lambda}\Gamma^{\lambda}_{\nu\alpha} + \frac{1}{16}(\omega_{\mu a b})^2$&$\frac{1}{8}g^{\mu\nu}\Gamma^{\alpha}_{\mu\lambda}\Gamma^{\lambda}_{\nu\alpha} + \frac{1}{8}(\omega_{\mu a b})^2$\\
\cline{3-4}
& &$\frac{1}{16}g^{\mu\nu}\Gamma^{\alpha}_{\mu\lambda}\Gamma^{\lambda}_{\nu\alpha}$&$0$\\
\hline
DR &$-\frac{1}{8}R$&$0$&$0$\\
\cline{3-4}
& &$0$&$0$\\
\hline
\end{tabular}
\end{table}

It is well known that Feynman diagrams lead to different results depending on the scheme chosen,
but $\mathcal{Z}(\beta)$ has to be unique, therefore scheme-dependent counterterms are needed to
recover the physical result. This is the general philosophy of renormalizable QFT, and quantum
mechanics can be considered as a particular QFT which lives in $D=0+1$ dimensions. Power counting
considerations show that the model is super-renormalizable, so that a two-loop computation,
\emph{i.e.} up to first order in $\beta$ (as $\beta^{L-1}$ indicates the loop dependence of the
correction), is sufficient to fix the counterterms. The known counterterms related to the different
regularization schemes are listed in Table~\ref{table:preliminar} (for the $N=0$ case we take as
quantum Hamiltonian $H=-\nabla^2/2$ without non-minimal coupling to the scalar curvature).

Mode Regularization for bosonic ($N=0$) nonlinear sigma models was studied and used for trace
anomalies calculations in~\cite{Bastianelli:1991be,Bastianelli:1992ct} and the complete counterterm
was obtained in~\cite{Bastianelli:1998jm}. In this paper we study the extension of MR to $N=1,2$
supersymmetric sigma models.

Time Slicing is the natural regularization that arises in the derivation of the path integral from the operatorial methods using the relation between Weyl ordering and the midpoint prescription~\cite{Berezin:1971jf}. Weyl ordering of the quantum Hamiltonian was used in~\cite{Mizrahi:1975pw} to identify the bosonic counterterm. Its non covariant part was derived independently in~\cite{Gervais:1976ws} performing a change of coordinates (point canonical transformations) in the Hamiltonian in flat space. By carefully
studying the relation between operator methods, discretized and continuous path integrals, the Feynman rules to be used in the continuum limit for Time Slicing were derived in~\cite{deBoer:1995hv}, while the extensions to $N=1,2$ can be found in~\cite{DeBoer:1995cb}, where the $N=1,2$ TS counterterms are derived by Weyl ordering the supersymmetric Hamiltonians.

Dimensional Regularization was applied to quantum mechanics in~\cite{Kleinert:1999aq}, where the
absence of non-covariant counterterms was noted. The complete counterterm was found
in~\cite{Bastianelli:2000pt,Bastianelli:2000nm}. The extensions to $N=1$ and $N=2$ were studied
in~\cite{Bastianelli:2002qw} and~\cite{Bastianelli:2005vk}, respectively. An extensive discussion
of these regularization schemes can be found in~\cite{Bastianelli:book}.

Before describing our calculation, let us recall that the regularization scheme also includes a
treatment of the functional measures; the bosonic one is suitably covariantized:
\begin{equation*}
  \mathcal{D}x \sim \prod_\tau \sqrt{g}Dx \sim \prod_\tau \sqrt{g(x(\tau))}\de^D x(\tau) \quad ;
\end{equation*}
while the covariant fermionic measure is the standard functional measure $\prod_{\tau} \de^D
\psi_i(\tau)$ if worldline fermions are chosen to carry flat indices. In order to obtain
translational invariant measures, useful for perturbative calculations, we rewrite
$\prod_{\tau}\sqrt{g}$ by a path integral over auxiliary ghost
fields~\cite{Bastianelli:1991be,Bastianelli:1992ct}:
\begin{equation*}
  \prod_\tau \sqrt{g(x(\tau))} \propto \int Da Db Dc \; e^{-S_{gh}} \quad ,
\end{equation*}
where
\begin{equation*}
  S_{gh}= \frac{1}{\beta}\int_{-1}^0 \ide{\tau}\frac{1}{2}g_{\mu\nu}(x)\left(a^\mu a^\nu + b^\mu c^\nu \right) \quad ;
\end{equation*}
with $a$ being a commuting field while $b$ and $c$ anticommuting. As we shall see, ghosts contributions cancel potential
infinities from Feynman diagrams, leaving a finite remainder; the comparison with the expected answer for the transition
amplitude (this takes the role of imposing the necessary renormalization conditions) fixes the counterterm.
In the following we will compute such counterterms for the susy sigma models with both flat and curved indices for
fermions. We start with the $N=1$ model, and then perform the $N=2$ calculations which are quite similar.

\section{N=1 Sigma Model}
\label{sec:N=!}
\subsection{Flat Indices}
\label{sec:flat-indices}

The total quantum action for the $N=1$ susy sigma model is
\begin{equation}
  \label{eq:S-tot}
  S = \frac{1}{\beta}\int_{-1}^0 \ide{t}\Big[\,\frac{1}{2}\,g_{\mu\nu}(x)\,\Big(\dot{x}^\mu\dot{x}^\nu+a^\mu a^\nu+b^\mu c^\nu\Big)+\frac{1}{2}\,\psi^a\Big(\dot{\psi}_a+\omega_{\mu\,ab}(x)\dot{x}^\mu\psi^b\Big)+\beta^2V_{MR}(x)\Big]\; ;
\end{equation}
where $V_{MR}$ is the mode-regularization counterterm we have to find, necessary to make contact
with transition amplitudes calculated from $\widehat{H}=-\slashed{\nabla}^2/2$.

In order to fix $V_{MR}$ we perform the two loop calculation of
\begin{equation}
  \label{eq:trans:element}
  \mathcal{K}(x_0,\beta) = \tr \Braket{x_0}{x_0}{e^{-\beta \widehat{H}}} = \int\limits_{x(-1)=x(0)=x_0}\mspace{-28mu}DxDaDbDc \int\nolimits_{ABC}\mspace{-10mu}D\psi \; e^{-S} \quad ,
\end{equation}
where the trace is performed only over $\psi$'s; and we compare it with the same transition element
calculated with the other regularization schemes. First of all we split the action in free ($S_2$)
and interacting ($S_{int}$) parts, \emph{i.e.}
\begin{align}
  \label{eq:S_2}
  S_2 = &\frac{1}{\beta}\int_{-1}^0\ide{\tau}\Big[\,\frac{1}{2}\,g_{\mu\nu}(x_0)\,\Big(\dot{x}^\mu\dot{x}^\nu+a^\mu a^\nu+b^\mu c^\nu\Big)+\frac{1}{2}\,\psi^a\dot{\psi}_a\Big]\; ;\\
 S_{int} = &\frac{1}{\beta}\int_{-1}^0\ide{\tau}\Big[\,\frac{1}{2}\,\Big(\,g_{\mu\nu}(x)-g_{\mu\nu}(x_0)\Big)\,\Big(\dot{x}^\mu\dot{x}^\nu+a^\mu a^\nu+b^\mu c^\nu\Big) + \frac{1}{2}\,\omega_{\mu\,ab}(x)\,\dot{x}^\mu\psi^a\psi^b\notag\\
\label{eq:S_int}
&+\beta^2\,V_{MR}(x)\Big]\; ;
\end{align}
so denoting as usual the normalized free average of a function $f$ with $\ave{f}$, $\mathcal{K}(x_0,\beta)$ up to order $\beta$ reduces to
\begin{equation*}
  \mathcal{K}(x_0,\beta) = A \ave{e^{-S_{int}}} = A\left[ 1-\ave{S_3} +\frac{1}{2}\ave{S_3^2} - \ave{S_4} \right] \quad ,
\end{equation*}
where $\ave{S_k}$ is the part of the action of order $\beta^{k/2-1}$, and $A$ is the value of the free path integral,
\begin{equation*}
  A=\int\limits_{x_0\to x_0}\mspace{-10mu}DxDaDbDc \int\nolimits_{ABC}\mspace{-10mu}D\psi \; e^{-S_2} = (\pi\beta)^{-D/2} \quad .
\end{equation*}

We now perform the usual classical background - quantum fluctuations split:
$x^{\mu}(\tau)=x_0^\mu+q^\mu(\tau)$ , where $q^\mu(-1)=q^\mu(0)=0$, (vanishing boundary
conditions); so we can write $S_3$ and $S_4$ as:
\begin{gather}
  \label{eq:S_3}
  S_3 = \frac{1}{\beta}\int_{-1}^0\ide{\tau}\Big[\,\frac{1}{2}\,\partial_\lambda
g_{\mu\nu}\,\Big(q^\lambda\dot{q}^\mu\dot{q}^\nu+q^\lambda a^\mu a^\nu+q^\lambda b^\mu
c^\nu\Big)+\frac{1}{2}\,\omega_{\mu\,ab}\,\dot{q}^\mu\psi^a\psi^b\Big]\, ,\\
\label{eq:S_4}
S_4 = \frac{1}{\beta}\int_{-1}^0\ide{\tau}\Big[\,\frac{1}{4}\,\partial_\lambda\partial_\sigma g_{\mu\nu}\,q^\lambda q^\sigma\Big(\dot{q}^\mu\dot{q}^\nu+a^\mu a^\nu+b^\mu c^\nu\Big)+\frac{1}{2}\,\partial_\lambda\omega_{\mu\,ab}\,q^\lambda\dot{q}^\mu\psi^a\psi^b\Big] +\beta\,V_{MR}\; ;
\end{gather}
from now on all the $x$-dependent functions are intended to be calculated at the point $x_0$ if not otherwise specified.
According to the vanishing boundary conditions\footnote{ghosts have the same boundary conditions as $q$'s.}, we expand
the $q,a,b,c$ fields in a sine series, obtaining
\begin{equation*}
  \phi^\mu(\tau) = \sum_{m=1}^{\infty}\phi_m^\mu \sin(\pi m\tau) \quad ,
\end{equation*}
where $\phi$ stands for one of the already mentioned fields. On the other side $\psi$'s have antiperiodic boundary
conditions, so we expand these fields with half-integer modes ($r=\pm 1/2,\pm 3/2,\dotsc$):
\begin{equation*}
  \psi^a(\tau) = \sum_{r\,\in\,\mathbb{Z}+1/2}\psi^a_r \; e^{2\pi i r \tau} \quad .
\end{equation*}
We perform mode regularization by introducing an integer mode cut-off $M$, so that the infinite
sums become:\begin{equation*}
  \sum_{m=1}^{\infty} \to \sum_{m=1}^M \quad , \quad \sum_{r\,\in\,\mathbb{Z}+1/2} \to \sum_{r=-M -1/2}^{M+1/2} \quad ;
\end{equation*}
so that we can define the regulated functional measure as
\begin{equation*}
  DqDaDbDcD\psi \propto \lim_{M\to\infty} \prod_{m=1}^M \prod_{a=1}^D\prod_{r=1/2}^{M+1/2}\de^Dq_m\de^Da_m\de^Db_m\de^Dc_m\de\psi^a_{-r}\de\psi^a_r \quad .
\end{equation*}

Performing the $\tau$-integral in $S_2$, introducing sources and completing squares as usual we obtain the following
two-point correlation functions or propagators, all the others being zero:
\begin{align*}
\ave{q^\mu(\tau)q^\nu(\sigma)}&=-\beta g^{\mu\nu}(x_0)\Delta(\tau,\sigma)\quad ,\\
\ave{a^\mu(\tau)a^\nu(\sigma)}&=\beta g^{\mu\nu}(x_0)\Delta_{gh}(\tau,\sigma)\quad,\\
\ave{b^\mu(\tau)c^\nu(\sigma)}&=-2\beta g^{\mu\nu}(x_0)\Delta_{gh}(\tau,\sigma)\quad,\\
\ave{\psi^a(\tau)\psi^b(\sigma)}&=\beta\delta^{ab}\Delta_{AF}(\tau-\sigma)\quad;
\end{align*}
with\footnote{the subscripts $_{gh}$ and $_{AF}$ stand for ghosts and Antiperiodic Fermions respectively.}
\begin{align*}
\Delta(\tau,\sigma)=-\sum_{m=1}^M\frac{2}{\pi^2m^2}\sin(\pi m\tau)\sin(\pi m \sigma) &\xrightarrow{M\to\infty} \tau(\sigma+1)\theta(\tau-\sigma) + \sigma(\tau+1)\theta(\sigma -\tau)\quad ,\\
\Delta_{gh}(\tau,\sigma)=2\sum_{m=1}^M\sin(\pi m\tau)\sin(\pi m \sigma)&\xrightarrow{M\to\infty}\delta(\tau,\sigma)\quad,\\
\Delta_{AF}(\tau-\sigma)=\sum_{r=-M-1/2}^{M+1/2}\frac{1}{2\pi ir}\,e^{2\pi ir(\tau-\sigma)}&\xrightarrow{M\to\infty}\frac{1}{2}\epsilon(\tau-\sigma) \quad ,
\end{align*}
where $\delta(\tau,\sigma)$ and $\theta(\tau-\sigma)$ act on functions with compact support on
$[-1,0]$ while $\epsilon(\tau-\sigma)$ is the sign distribution acting on antiperiodic functions.

We are now ready to make perturbative calculations on $S_3$, $S_3^2$ and $S_4$ using standard Wick contractions and the propagators listed above, obtaining:
\begin{align}
\label{eq:ave_S3}
\ave{S_3} =& 0\quad ,\\
\label{eq:ave_S4}
\ave{S_4} =& \frac{\beta}{4}\,\partial_\lambda\partial_\sigma g_{\mu\nu}\,\Big[g^{\lambda\sigma}g^{\mu\nu}\mathbf{I}_1+2g^{\lambda\mu}g^{\sigma\nu}\mathbf{I}_2\Big]-\frac{\beta}{2}\,\partial_\sigma\omega_{\mu\,ab}g^{\sigma\mu}\delta^{ab}\mathbf{I}_3+\beta\,V_{MR}\quad ,\\
\ave{S_3^{\,2}} =& -\frac{\beta}{4}\,\partial_\lambda g_{\mu\nu}\,\partial_\sigma g_{\alpha\beta}\,\Big[g^{\lambda\sigma}g^{\alpha\beta}g^{\mu\nu}\mathbf{I}_4+2g^{\lambda\sigma}g^{\mu\alpha}g^{\nu\beta}\mathbf{I}_5+ 4g^{\lambda\mu}g^{\nu\sigma}g^{\alpha\beta}\mathbf{I}_6\quad\phantom{,}\notag\\
&+4g^{\lambda\alpha}g^{\mu\sigma}g^{\nu\beta}\mathbf{I}_7+4g^{\lambda\mu}g^{\sigma\alpha}g^{\nu\beta}\mathbf{I}_8\Big]+ \frac{\beta}{2}\,\partial_\lambda g_{\mu\nu}\,\omega_{\sigma\,ab}\,\Big[g^{\lambda\sigma}g^{\mu\nu}\delta^{ab}\mathbf{I}_9\quad\phantom{,}\notag\\
\label{eq:ave_S3sq}
&+2g^{\lambda\mu}g^{\sigma\nu}\delta^{ab}\mathbf{I}_{10}\Big]-\frac{\beta}{4}\,\omega_{\mu\,ab}\,\omega_{\nu\,cd}\,\Big[g^{\mu\nu}\delta^{ab}\delta^{cd}\mathbf{I}_{11}-2g^{\mu\nu}\delta^{ac}\delta^{bd}\mathbf{I}_{12}\Big]\quad ,
\end{align}
where the $\mathbf{I}_k$ are listed in Appendix~\ref{cha:feynman-diagrams}. The value of $\mathbf{I}_{12}$ is indicated with $\mathcal{I}$ because we were not able to compute it analytically (though numerically it is seen to converge to $1/6$). These contributions sum up, at order $\beta$, to:
\begin{equation}
 \label{eq:Kvbc}
 \mathcal{K}(x_0,\beta)= \frac{1}{(\pi\beta)^{D/2}}\left[ 1-\beta\left( \frac{1}{24}R +\frac{1}{24}g_{\gamma\sigma}g^{\alpha\mu}g^{\beta\nu}\Gamma^{\gamma}_{\phantom{\gamma}\alpha\beta}\Gamma^{\sigma}_{\phantom{\sigma}\mu\nu} - \frac{\mathcal{I}}{4}g^{\mu\nu}\omega_{\mu\,ab}\omega_{\nu}^{\phantom{\nu}ab} + V_{MR}\right) \right] \; .
\end{equation}
This result can be compared with the one obtained employing other regularization
schemes \cite{DeBoer:1995cb,Bastianelli:2002qw} that reads
\begin{equation}
\label{eq:Vdr}
  \mathcal{K} = \frac{1}{(\pi\beta)^{D/2}}\left[1-\frac{\beta}{24}R + O(\beta^2)\right] \quad .
\end{equation}
Hence we obtain
\begin{equation}
\label{eq:Vmr_unfixed}
  V_{MR} = - \frac{1}{24}\Gamma^2 + \frac{\mathcal{I}}{4}\omega^2 \quad ,
\end{equation}
with the index contraction rules given in~\eqref{eq:Kvbc}. The part independent on $\mathcal{I}$ is
part of the bosonic MR counterterm\footnote{The covariant piece $-R/8$ is not subtracted because
its presence is demanded by the quantum Hamiltonian.} (see
table~\ref{table:preliminar})~\cite{Bastianelli:1998jm}, while the remainder is due to the
fermions.

In order to fix $\mathcal{I}$ analytically we calculate the partition function
\begin{equation*}
  Z[\beta]= \int \ide{^D x}\sqrt{g(x)}\mathcal{K}(x,\beta) = \Tr e^{-\beta\widehat{H}} = \int_{PBC}\mspace{-16mu}DxDaDbDc\int_{ABC}\mspace{-16mu}D\psi \: e^{-S} \; .
\end{equation*}
In fact, periodic boundary conditions permit the use of translational invariant propagators (the so-called string
inspired propagators) which are simpler to deal with; the drawback is that since $\mathcal{K}(x,\beta)$ is integrated
over $x$, we loose information about total derivatives that could affect $V_{MR}$, but since
Eq.~(\ref{eq:Vmr_unfixed}) shows that this counterterm does not contain such terms, we do not have to care about them
(see discussions in~\cite{Schalm:1998ix,Bastianelli:2003bg}).

First of all we expand $x,a,b,c$ respecting periodic boundary conditions, in particular we separate the $x$ zero mode
from the rest:
\begin{equation*}
  x^\mu(\tau) = \sum_{-M}^{M} q^\mu_m \, e^{2\pi im\tau} = x_0^\mu + \sum_{\substack{m=-M\\m\neq 0}}^M q^\mu_m \, e^{2\pi im\tau} = x_0^\mu + q^\mu(\tau) \quad ;
\end{equation*}
and
\begin{equation*}
  a^\mu(\tau) = \sum_{-M}^{M} a^\mu_m \, e^{2\pi im\tau} \quad , \quad b^\mu(\tau) = \sum_{-M}^{M} b^\mu_m \, e^{2\pi im\tau} \quad , \quad c^\mu(\tau) = \sum_{-M}^{M} c^\mu_m \, e^{2\pi im\tau} \quad ;
\end{equation*}
while $\psi$'s are expanded as before. So the measure splits into:
\begin{equation*}
  DxD\psi DaDbDc = \de^Dx_0DqD\psi DaDbDc \quad .
\end{equation*}
Finally we can write $Z[\beta]$ as:
\begin{equation*}
  Z[\beta] = \int \de^D x_0 A(x_0,\beta) \ave{e^{-S_{int}}} \quad ,
\end{equation*}
where $A$ contains an extra $\sqrt{g(x_0)}$ factor due to the ghost's zero mode, \emph{i.e.}
\begin{equation*}
A=(\pi\beta)^{-D/2}\sqrt{g(x_0)}\; .
\end{equation*}
The propagators now become
\begin{align*}
\ave{q^\mu(\tau)q^\nu(\sigma)}&=-\beta\,g^{\mu\nu}(x_0)\,\Delta_{SI}(\tau-\sigma)\quad ,\\
\ave{a^\mu(\tau)a^\nu(\sigma)}&=\beta\,g^{\mu\nu}(x_0)\,\Delta_{GH}(\tau-\sigma)\quad ,\\
\ave{b^\mu(\tau)c^\nu(\sigma)}&=-2\beta\,g^{\mu\nu}(x_0)\,\Delta_{GH}(\tau-\sigma) \quad ,
\end{align*}
where\footnote{$_{SI}$ denotes string inspired propagators, while $_{GH}$ denotes ghost propagator with periodic boundary conditions.}
\begin{align*}
  \Delta_{SI}(x)&=-\sum_{\substack{m=-M\\m\neq0}}^M\frac{1}{4\pi^2m^2}\,e^{2\pi imx}\xrightarrow{M\to\infty}-\frac{1}{2}x^2 +\frac{1}{2}\ass{x}-\frac{1}{12} \quad, \quad x \in [-1,1] \quad ,\\
\Delta_{GH}(x)&=\sum_{m=-M}^Me^{2\pi imx} \xrightarrow{M\to\infty}\delta(x)\quad.
\end{align*}
Fermionic propagators are the same as before. The structure of $\ave{S_3}$, $\ave{S_4}$,
$\ave{S_3^2}$ is the same as in Eq.s~(\ref{eq:ave_S3},~\ref{eq:ave_S4},~\ref{eq:ave_S3sq}), but the
$\mathbf{I}_k$ take now different values, as reported in Appendix~\ref{cha:feynman-diagrams}. Using
the counterterm in (\ref{eq:Vmr_unfixed}), which depend explicitly on $\mathcal{I}$, we obtain
\begin{equation}
\label{eq:Z_final}
  \begin{split}Z[\beta] = \frac{1}{(\pi\beta)^{D/2}}\int\ide{^Dx_0}\sqrt{g(x_0)} \Bigg[ 1+ \beta\left( -\frac{1}{24}R + \frac{1}{24}\omega^2 - \frac{\mathcal{I}}{4}\omega^2 + \frac{1}{\sqrt{g(x_0)}}\partial_\mu\mathcal{A}^\mu\right) \\+ O(\beta^2)\Bigg]\end{split}
\end{equation}
where the total derivative $\partial_\mu\mathcal{A}^\mu$ can be dropped; however for the sake of completeness we write
$\mathcal{A}^\mu$ explicitly:
\begin{equation*}
  \mathcal{A}^\mu = \sqrt{g} \left( \frac{1}{24}g^{\mu\nu}g^{\alpha\beta}\partial_\alpha g_{\beta\nu} - \frac{1}{48} g^{\mu\nu}g^{\alpha\beta}\partial_\nu g_{\alpha\beta} \right) \quad .
\end{equation*}

The partition function in (\ref{eq:Z_final}) is consistent with the result in (\ref{eq:Vdr}) if
$\mathcal{I}=1/6$, so the counterterm $V_{MR}$ is given by
\begin{equation}
  \label{eq:Vmr_flat}
  V_{MR} = -\frac{1}{24} g_{\gamma\sigma}g^{\alpha\mu}g^{\beta\nu}\Gamma^{\gamma}_{\phantom{\gamma}\alpha\beta}\Gamma^{\sigma}_{\phantom{\sigma}\mu\nu} + \frac{1}{24}g^{\mu\nu}\omega_{\mu\,ab}\omega_{\nu}^{\phantom{\nu}ab} \quad .
\end{equation}

\subsection{Curved Indices}
\label{sec:curved-indices}

The result just found for $V_{MR}$ suggests that a more symmetric treatment of the superpartners
$x$ and $\psi$ can make the counterterm vanish by supersymmetry, as we will see. For this reason,
we introduce worldline fermions with curved indices contracting the $\psi$'s with the vielbein:
\begin{equation*}
  e^\mu_a(x)\psi^a \equiv \psi^\mu \quad .
\end{equation*}
Using such new fields as dynamical variables, the susy sigma model action becomes
\begin{equation*}
  S= \frac{1}{\beta}\int_{-1}^0\de\tau\,\Big\{\,\frac{1}{2}g_{\mu\nu}(x)\Big[\dot{x}^\mu\dot{x}^\nu+\psi^\mu\dot{\psi}^\nu+ \psi^\mu\Gamma^\nu_{\phantom{\nu}\alpha\lambda}(x)\psi^\lambda \dot{x}^\alpha\Big]+\beta^2V'_{MR}(x)\Big\}\quad ,
\end{equation*}
where $V'_{MR}$ is the quantum counterterm. It is worthwhile noting that now space-time gravity is
described only by means of the metric tensor and Christoffel coefficients. This is a nice feature,
since the 1-D susy sigma model can be used for doing 1-loop calculations in the QFT of a Dirac
field (see, for example,~\cite{Bastianelli:2002qw}), and space-time fermions are coupled to gravity mainly
through the vielbein formalism. Using antisymmetry of Grassmann variables the action simplifies to
\begin{equation*}
  S=\frac{1}{2\beta}\int_{-1}^0\de\tau\Big[\,g_{\mu\nu}\dot{x}^\mu\dot{x}^\nu+g_{\mu\nu}\psi^\mu\dot{\psi}^\nu-\partial_\mu g_{\nu\alpha}\psi^\mu\psi^\nu\dot{x}^\alpha+2\beta^2V'_{MR}\Big] \quad .
\end{equation*}
Since fermions carry curved vector indices, their covariant measure is defined accordingly as\footnote{$g$ is to the power $-1/2$ since fermionic fields are Grassmann variables.}
\begin{equation*}
  \mathcal{D}\psi = \prod_\tau \frac{1}{\sqrt{g(x(\tau))}}D\psi \quad .
\end{equation*}
The procedure to fix the counterterm is perfectly analogous to the flat indices case of
Section~\ref{sec:flat-indices}, but to manage the $1/\sqrt{g}$ factors in the measure, we introduce
a new commuting ghost field $\alpha^\mu(\tau)$ as in~\cite{Bastianelli:2002qw}, so the total action
$S[x,a,b,c,\psi,\alpha]$ reads:
\begin{equation}
  \label{eq:S_curved}
  S=\frac{1}{2\beta}\int_{-1}^0\de\tau \Big[\,g_{\mu\nu}\left(\dot{x}^\mu\dot{x}^\nu+\psi^\mu\dot{\psi}^\nu + a^\mu a^\nu+b^\mu c^\nu+\alpha^\mu\alpha^\nu\right)-\partial_\mu g_{\nu\alpha}\psi^\mu\psi^\nu\dot{x}^\alpha+2\beta^2V'_{MR}\Big] \quad.
\end{equation}
The free part $S_2$ now results:
\begin{equation*}
  S_2 = \frac{g_{\mu\nu}(x_0)}{2\beta}\int_{-1}^0\ide{\tau} \left[ \dot{x}^\mu\dot{x}^\nu + \psi^\mu\dot{\psi}^\nu+a^\mu a^\nu + b^\mu c^\nu + \alpha^\mu\alpha^\nu \right] \quad .
\end{equation*}
The $\psi$ propagator is slightly modified and reads
\begin{equation*}
  \ave{\psi^\mu(\tau)\psi^\nu(\sigma)} = \beta g^{\mu\nu}(x_0) \Delta_{AF}(\tau-\sigma) \quad ;
\end{equation*}
and since $\alpha$'s are related to $\psi$'s they have anti-periodic boundary conditions as well.
Expanding them accordingly and finding propagators in the usual manner we find\footnote{$_{FG}$ stands for Fermion's Ghosts.}:
\begin{equation*}
  \ave{\alpha^\mu(\tau)\alpha^\nu(\sigma)} = \beta g^{\mu\nu}(x_0) \Delta_{FG}(\tau -\sigma) \: , \quad \text{where} \quad \Delta_{FG}(x)=\mspace{-15mu}\sum_{r=-M-1/2}^{M+1/2}e^{2\pi irx}\xrightarrow{M\to\infty}\delta_A(x) \; ,
\end{equation*}
where $x\in [-1,1]$ and $\delta_A$ is the delta distribution acting on anti-periodic functions; while the other propagators remain the same as in Section~\ref{sec:flat-indices}.

The interacting action up to order $\beta$ is $S_{int}=S_3+S_4$, where\footnote{the $x$ field has already been split in classical background - quantum fluctuation.}
\begin{align*}
  S_3=&\,\frac{1}{2\beta}\,\partial_\alpha g_{\mu\nu}\int_{-1}^0\de\tau\,\Big[q^\alpha\dot{q}^\mu\dot{q}^\nu+q^\alpha\psi^\mu\dot{\psi}^\nu+q^\alpha(a^\mu a^\nu+b^\mu c^\nu)+q^\alpha\alpha^\mu\alpha^\nu+\dot{q}^\mu\psi^\nu\psi^\alpha\Big] \; ,\\
S_4=&\,\frac{1}{\beta}\int_{-1}^0\de\tau\Big[\,\frac{1}{4}\partial_\alpha\partial_\beta g_{\mu\nu}q^\alpha q^\beta\Big(\dot{q}^\mu\dot{q}^\nu+\psi^\mu\dot{\psi}^\nu+a^\mu a^\nu+b^\mu c^\nu+\alpha^\mu\alpha^\nu\Big)\\
&+\frac{1}{2}\partial_\beta\partial_\mu g_{\nu\alpha}\psi^\nu\psi^\mu\dot{q}^\alpha q^\beta\Big]+\beta V'_{MR} \; .
\end{align*}

For the sake of simplicity we introduce a condensed notation as follows:
\begin{align*}
(\partial_\alpha g_{\mu\nu})^2&=g^{\alpha\beta}\,g^{\mu\lambda}\,g^{\nu\sigma}\,\partial_\alpha g_{\mu\nu}\,\partial_\beta\, g_{\lambda\sigma}\\
(\partial_\alpha g_{\mu\nu})\,(\partial_\mu g_{\alpha\nu})&=g^{\alpha\beta}\,g^{\mu\lambda}\,g^{\nu\sigma}\,\partial_\alpha g_{\mu\nu}\,\partial_\lambda\, g_{\beta\sigma}\\
\partial_\beta g= g^{\mu\nu}\,\partial_\beta g_{\mu\nu}\;,\quad g_\beta&= g^{\mu\nu}\,\partial_\mu g_{\beta\nu}\;,\quad g^\beta=g^{\beta\mu}\,g_\mu\\
\partial^2g=g^{\alpha\beta}\,g^{\mu\nu}\,\partial_\alpha\partial_\beta g_{\mu\nu}\;&,\quad\partial^\alpha g_\alpha=g^{\alpha\beta}\,g^{\mu\nu}\,\partial_\alpha\partial_\mu g_{\beta\nu}\;.
\end{align*}
With this notation the averages of $S_3$, $S_4$ and $S_3^2$ result:
\begin{align*}
\ave{S_3} &=\,0 \quad ,\\
\ave{S_4} &= \,\frac{\beta}{4}\left[ \partial^2g(\mi{1}-\mi{13})+2\partial_\alpha g^\alpha (\mi{2}+\mi{3}) \right] +\beta V'_{MR}\quad ,\\
\ave{S_3^2} &= -\frac{\beta}{4}\left[ \partial_\alpha g\partial^\alpha g (\mi{4}-2\mi{15}+\mi{16}-2\mi{17}) + (\partial_\alpha g_{\mu\nu})^2\left( 2\mi{5}-\mi{12}-\mi{19} \right) \right.\\
&\mspace{-5mu}\left.+g^\alpha\partial_\alpha g(4\mi{6}-2\mi{9}-4\mi{14}+2\mi{18}) + \partial_\alpha g_{\mu\nu}\partial_\mu g_{\alpha\nu}(4\mi{7}+\mi{12}) + g_\alpha g^\alpha (4\mi{8}-4\mi{10}+\mi{11})\right] \, ;
\end{align*}
the $\mi{k}$ are again reported in Appendix~\ref{cha:feynman-diagrams}. Hence, summing up and
comparing with Eq.~\eqref{eq:Vdr}, we fix the counterterm in the case of curved indices:
\begin{equation}
\label{eq:V_mr:curved}
  \mathcal{K}(x_0,\beta)= \frac{1}{(\pi\beta)^{D/2}}\left[1-\beta(\frac{1}{24}R+V'_{MR})+O(\beta^2)\right] \quad \Rightarrow \quad V'_{MR}= 0 \quad .
\end{equation}
We see, as anticipated at the beginning, that fermionic and bosonic contributions to the
counterterm are equal in magnitude and cancel out (while in DR they are separately zero), leaving a
covariant and supersymmetric action. The price of introducing new ghost variables and different fermionic
vertices, actually make perturbative calculations slightly more efficient.

\section{N=2 Sigma Model}
\label{sec:n=2-sigma-model}

Extending the supersymmetric partners of the $x$ fields to the doublet $\psi_i$ ($i=1,2$) with
$O(2)$ internal symmetry we obtain the sigma model with $N=2$ extended supersymmetries, whose
actions, in the cases of flat and curved indices, read
\begin{gather}
\label{eq:total-S-susy-N=2-flat}
 \begin{split}S_f[x,\psi_1,\psi_2]=\frac{1}{\beta}\int_{-1}^0 \ide{\tau} \Bigl[ \frac{1}{2}g_{\mu\nu}(x)\Bigl(\dot{x}^\mu\dot{x}^\nu + a^\mu a^\nu + b^\mu c^\nu\Bigr) + \frac{1}{2}\psi_{ai}\dot{\psi}^a_i \\
+ \frac{1}{2}\omega_{\mu a b}(x)\dot{x}^\mu\psi_i^a\psi_i^b - \frac{1}{8}R_{abcd}(x)\psi_i^a\,\psi_i^b\,\psi_k^c\,\psi_k^d + \beta^2V_{MR}(x) \Bigr]\quad ,\end{split}\\
\label{eq:total-S-susy-N=2-curved}
\begin{split} S_c[x,\psi_1,\psi_2] = \frac{1}{\beta}\int_{-1}^0\ide{\tau} \Bigl[ \frac{1}{2}g_{\mu\nu}(x)\Bigl(\dot{x}^\mu \dot{x}^\nu + \psi_i^\mu \dot{\psi}_i^{\nu} + a^\mu a^\nu + b^\mu c^\nu + \alpha_i^\mu\alpha_i^\nu\Bigr) \\
+ \frac{1}{2}g_{\mu\nu}(x)\psi^\mu_i \dot{x}^\lambda
\Gamma^\nu_{\phantom{\nu}\lambda\sigma}(x)\psi_i^{\sigma} -
\frac{1}{8}R_{\mu\nu\lambda\sigma}(x)\psi_i^\mu\,\psi_i^\nu\,\psi_k^\lambda\,\psi_k^\sigma +
\beta^2 V'_{MR} \Bigr] \quad .\end{split}
\end{gather}
These actions allow to compute amplitudes with Hamiltonian\footnote{$Q_i$, $i=1,2$ are the
conserved supercharges}:
\begin{equation*}
\widehat{H}=-\frac{1}{2}\nabla^2-\frac{1}{8}R_{abcd}\psi^a_i\psi^b_i\psi^c_k\psi^d_k=\frac{1}{4}\big\{Q_i\,,\,Q_i\big\} \; ;
\end{equation*}
and, as in the previous section, we will derive the required counterterm by a two-loop calculation
of $\tr\Braket{x_0}{x_0}{e^{-\beta\widehat{H}}}$, where the trace is taken over the fermionic
Hilbert space only. The introduction of ghosts is perfectly analogous to the $N=1$ case. The
propagators are the same as before, and diagonal in fermion species
\begin{equation*}
  \ave{\psi_i^{a(\mu)}(\tau)\psi_j^{b(\nu)}(\sigma)}=\beta\delta^{ab}(g^{\mu\nu})\delta_{ij}\Delta_{AF}(\tau-\sigma) \quad .
\end{equation*}
The term proportional to $R\psi\psi\psi\psi$ gives a vanishing contribution at two-loop level. The
fermionic part of $S_3$ splits into $S_{3,1}+S_{3,2}$, each part depending only on a single
fermionic specie, and furthermore the mixed part $2S_{3,1}S_{3,2}$ gives a null contribution to
$\ave{S_3^2}$. For this reason, the fermionic contribution to $V_{MR}$ is simply doubled with respect to the $N=1$ case; thus the cancellation between bosonic and fermionic terms does not occur, leaving unfortunately a non-covariant quantum action. Performing calculations, and comparing $\ave{e^{-S_{int}}}$
with the result given in~\cite{Bastianelli:2005vk}, \emph{i.e.} $1-\beta R/24$, we easily find the mode regularization counterterms for the $N=2$ model:
\begin{align*}
  V_{MR} &= \frac{1}{12}g^{\mu\nu}\omega_{\mu a b}\omega_{\nu}^{\phantom{\nu}ab}-\frac{1}{24}g_{\mu\alpha}g^{\nu\gamma}g^{\sigma\delta}\Gamma^{\mu}_{\phantom{\mu}\nu\sigma}\Gamma^{\alpha}_{\phantom{\alpha}\gamma\delta} \quad ,\\
V'_{MR} &= \frac{1}{24}g_{\mu\alpha}g^{\nu\gamma}g^{\sigma\delta}\Gamma^{\mu}_{\phantom{\mu}\nu\sigma}\Gamma^{\alpha}_{\phantom{\alpha}\gamma\delta} \quad .
\end{align*}

\section{Conclusions}
\label{sec:conclusions}

In this work we have completed the analysis of the known regularization schemes for the $N=1$ and $N=2$ nonlinear sigma
models by investigating MR. We have calculated the counterterm for the $N=1$ case using fermions with flat indices obtaining
$V_{MR}=-\Gamma^2/24+\omega^2/24$; the structure of such term suggested the possibility of compensation between
bosonic and fermionic parts in the case of curved indices. In fact an explicit calculation showed this to be the case:
the curved indices counterterm $V'_{MR}= 0$ vanishes leaving classical supersymmetry and covariance of the action
unbroken. \begin{table}[t]
\caption{Counterterms for $N=0,1,2$ sigma models.}
\label{table:two}
\begin{tabular}[c]{|c|c|c|c|}
\hline
 &$N=0$&$N=1$&$N=2$\\
\hline
MR &$-\frac{1}{8}R - \frac{1}{24}\left(\Gamma^{\mu}_{\nu\sigma}\right)^2$&$-\frac{1}{24}\left(\Gamma^{\mu}_{\nu\sigma}\right)^2+\frac{1}{24}(\omega_{\mu a b})^2$&$-\frac{1}{24}\left(\Gamma^{\mu}_{\nu\sigma}\right)^2+\frac{1}{12}(\omega_{\mu a b})^2$\\
\cline{3-4}
& &$0$&$\frac{1}{24}\left(\Gamma^{\mu}_{\nu\sigma}\right)^2$\\
\hline
TS &$-\frac{1}{8}R + \frac{1}{8}g^{\mu\nu}\Gamma^{\alpha}_{\mu\lambda}\Gamma^{\lambda}_{\nu\alpha}$&$\frac{1}{8}g^{\mu\nu}\Gamma^{\alpha}_{\mu\lambda}\Gamma^{\lambda}_{\nu\alpha} + \frac{1}{16}(\omega_{\mu a b})^2$&$\frac{1}{8}g^{\mu\nu}\Gamma^{\alpha}_{\mu\lambda}\Gamma^{\lambda}_{\nu\alpha} + \frac{1}{8}(\omega_{\mu a b})^2$\\
\cline{3-4}
& &$\frac{1}{16}g^{\mu\nu}\Gamma^{\alpha}_{\mu\lambda}\Gamma^{\lambda}_{\nu\alpha}$&$0$\\
\hline
DR &$-\frac{1}{8}R$&$0$&$0$\\
\cline{3-4}
& &$0$&$0$\\
\hline
\end{tabular}
\end{table}
Such compensation between bosonic and fermionic contributions is perhaps expected in supersymmetric models, although not necessary, since such terms depend on the regularization scheme chosen: in fact even if it holds also in dimensional regularization~\cite{Bastianelli:2002qw}, it is not true in time slicing. Furthermore we showed in section~\ref{sec:n=2-sigma-model} that for the $N=2$ model mode regularization gives a non-covariant, susy-breaking counterterm $\Gamma^2/24$, while both dimensional regularization and time slicing~\cite{Bastianelli:2005vk,DeBoer:1995cb} give a vanishing counterterm. In MR, one may interpret the vanishing of $N=1$ counterterm as due to the fact that this model is supersymmetric even off-shell (we have an equal number of bosonic and fermionic fields); and as a signal that such regularization scheme preserves the symmetry. On the other hand, in the $N=2$ model the number of fermionic and bosonic fields is not the same, and supersymmetry is realized only on-shell; this could be the reason for which a non-zero counterterm is needed at the quantum level to restore supersymmetry. Anyway direct calculation gives the explicit answer. With our finding we can summarize the counterterms for the various regularization schemes in Table \ref{table:two}.

\appendix

\section{Curvatures}
\label{sec:curvatures}

The vielbein field is related to the metric tensor via usual formula
\begin{equation*}
  g_{\mu\nu}(e(x))= \delta_{ab}e^a_\mu e^b_\nu \quad ;
\end{equation*}
and the vielbein postulate $\nabla_\mu e^a_\nu=0$ ensures the compatibility between Christoffel connection and metric, furthermore it relates $\Gamma$'s to $\omega$'s in the following way:
\begin{equation*}
  \omega_{\mu\phantom{a}b}^{\phantom{\mu}a}=e^a_{\nu}\partial_\mu e^\nu_b + e^a_\nu\Gamma^\nu_{\phantom{\nu}\mu\beta}e^\beta_b \quad .
\end{equation*}
Connection coefficients are given explicitly in terms of metric or vielbein by:
\begin{align*}
  \Gamma^\mu_{\phantom{\mu}\nu\sigma}&=\frac{1}{2}\,g^{\mu\lambda}\,\Big(\,\partial_\sigma g_{\lambda\nu}+\partial_\nu g_{\sigma\lambda}-\partial_\lambda g_{\nu\sigma}\Big)\;,\\
\omega_{\mu\,ab}&=\frac{1}{2}\,e^\nu_{\,a}\,(\,\partial_\mu e_{b\,\nu}-\partial_\nu e_{b\,\mu})-\frac{1}{2}\,e^\nu_{\,b}\,(\,\partial_\mu e_{a\,\nu}-\partial_\nu e_{a\,\mu})-\frac{1}{2}\,e_\mu^{\,c}\,e_a^{\,\nu}\,e_b^{\,\sigma}\,(\,\partial_\nu e_{c\,\sigma}-\partial_\sigma e_{c\,\nu}) \; .
\end{align*}

For the Riemann tensor we use the convention
\begin{equation*}
  \Big[\nabla_\mu,\,\nabla_\nu\Big]\,V^\lambda = R_{\mu\nu\phantom{\lambda}\sigma}^{\phantom{\mu\nu}\lambda}\,V^\sigma\;,
\end{equation*}
and we construct the Ricci tensor and the curvature scalar as:
\begin{equation*}
  R_{\mu\nu}=R_{\lambda\mu\phantom{\lambda}\nu}^{\phantom{\lambda\mu}\lambda}\;,\quad
R=R^\mu_{\phantom{\mu}\mu}\;>0\quad\text{on a sphere.}
\end{equation*}
Finally $R$ as a function of metric and its derivatives could be written as\footnote{using the condensed notation introduced above.}:
\begin{equation*}
  R = -\partial^{\,2}g+\partial^{\,\alpha} g_\alpha+\frac{3}{4}\,(\partial_\alpha
g_{\mu\nu})^2-\frac{1}{2}\,(\partial_\alpha g_{\mu\nu})\,(\partial_\mu g_{\alpha\nu})-\frac{1}{4}\,(\partial_\beta g)^2+(\partial_\beta g)\,g^\beta-g_\beta^2 \quad .
\end{equation*}

\section{Feynman Diagrams}
\label{cha:feynman-diagrams}

We report here the integrals $\mathbf{I}_k$ with their results and respective Feynman Diagrams. First of all we present the twelve integrals needed in the case of flat indices, with vanishing boundary conditions and for $N=1$:
\begin{align*}
\mathbf{I}_1 &= \;\parbox{16mm}{
    \begin{fmfgraph*}(16,8)
    \fmfset{wiggly_len}{1.5mm}\fmfset{dash_len}{1.5mm}
    \fmfipair{a,b,c,d,e}
    \fmfiequ{a}{.5[nw,sw]}
    \fmfiequ{b}{.5[ne,se]}
    \fmfiequ{c}{.5[a,b]}
    \fmfiequ{xpart(d)}{9mm}
    \fmfiequ{xpart(e)}{9mm}
    \fmfiequ{ypart(d)}{6.5mm}
    \fmfiequ{ypart(e)}{1.5mm}
    \fmfi{plain}{a..c..a}
    \fmfi{plain}{b..c..b}
    \fmfiv{d.sh=circle,d.siz=1.5thick}{d}
    \fmfiv{d.sh=circle,d.siz=1.5thick}{e}
    \end{fmfgraph*}}\;+\;
\parbox{16mm}{
    \begin{fmfgraph*}(16,8)
    \fmfset{wiggly_len}{1.5mm}\fmfset{dash_len}{1.5mm}
    \fmfipair{a,b,c}
    \fmfiequ{a}{.5[nw,sw]}
    \fmfiequ{b}{.5[ne,se]}
    \fmfiequ{c}{.5[a,b]}
    \fmfi{plain}{a..c..a}
    \fmfi{dashes}{b..c..b}
    \end{fmfgraph*}}\;=\int_{-1}^0\de\tau\Delta\rvert_\tau(\puntods{\Delta}+\Delta_{gh})\rvert_\tau=-\frac{1}{6}\; ,\\{}\\
\mathbf{I}_2&=\;\parbox{16mm}{
    \begin{fmfgraph*}(16,8)
    \fmfset{wiggly_len}{1.5mm}\fmfset{dash_len}{1.5mm}
    \fmfipair{a,b,c,d,e}
    \fmfiequ{a}{.5[nw,sw]}
    \fmfiequ{b}{.5[ne,se]}
    \fmfiequ{c}{.5[a,b]}
    \fmfiequ{xpart(d)}{7mm}
    \fmfiequ{xpart(e)}{9mm}
    \fmfiequ{ypart(d)}{6.5mm}
    \fmfiequ{ypart(e)}{1.5mm}
    \fmfi{plain}{a..c..a}
    \fmfi{plain}{b..c..b}
    \fmfi{plain}{a..c..a}
    \fmfiv{d.sh=circle,d.siz=1.5thick}{d}
    \fmfiv{d.sh=circle,d.siz=1.5thick}{e}
    \end{fmfgraph*}}\;=\int_{-1}^0\de\tau\puntos{\Delta}\rvert_\tau^2=\frac{1}{12}\; , \\{}\\
\mathbf{I}_3&=\;\parbox{16mm}{
    \begin{fmfgraph*}(16,8)
    \fmfset{wiggly_len}{1.5mm}\fmfset{dash_len}{1.5mm}
    \fmfipair{a,b,c,d}
    \fmfiequ{a}{.5[nw,sw]}
    \fmfiequ{b}{.5[ne,se]}
    \fmfiequ{c}{.5[a,b]}
    \fmfiequ{d}{1/4[nw,ne]}
    \fmfi{plain}{a..c..a}
    \fmfi{wiggly}{b..c..b}
    \fmfiv{d.sh=circle,d.siz=1.5thick}{d}
   \end{fmfgraph*}}\;=\int_{-1}^0\de\tau\puntos{\Delta}\rvert_\tau\Delta_{AF}\rvert_\tau=0\; , \\{}\\
\mathbf{I}_4&=\;\parbox{22mm}{
\begin{fmfgraph*}(22,8)
\fmfset{wiggly_len}{1.5mm}\fmfset{dash_len}{1.5mm}
    \fmfipair{a,b,c,d,e,f}
    \fmfiequ{a}{.5[nw,sw]}
    \fmfiequ{b}{.5[ne,se]}
    \fmfiequ{c}{4/11[a,b]}
    \fmfiequ{d}{7/11[a,b]}
    \fmfiequ{xpart(e)}{7mm}
    \fmfiequ{xpart(f)}{7mm}
    \fmfiequ{ypart(e)}{6.5mm}
    \fmfiequ{ypart(f)}{1.5mm}
    \fmfi{plain}{a..c..a}
    \fmfi{plain}{c--d}
    \fmfi{dashes}{b..d..b}
    \fmfiv{d.sh=circle,d.siz=1.5thick}{e}
    \fmfiv{d.sh=circle,d.siz=1.5thick}{f}
\end{fmfgraph*}}\;+\;
\parbox{22mm}{
\begin{fmfgraph*}(22,8)
\fmfset{wiggly_len}{1.5mm}\fmfset{dash_len}{1.5mm}
    \fmfipair{a,b,c,d,e,f}
    \fmfiequ{a}{.5[nw,sw]}
    \fmfiequ{b}{.5[ne,se]}
    \fmfiequ{c}{4/11[a,b]}
    \fmfiequ{d}{7/11[a,b]}
    \fmfiequ{xpart(e)}{15mm}
    \fmfiequ{xpart(f)}{15mm}
    \fmfiequ{ypart(e)}{6.5mm}
    \fmfiequ{ypart(f)}{1.5mm}
    \fmfi{plain}{b..d..b}
    \fmfi{plain}{c--d}
    \fmfi{dashes}{a..c..a}
    \fmfiv{d.sh=circle,d.siz=1.5thick}{e}
    \fmfiv{d.sh=circle,d.siz=1.5thick}{f}
\end{fmfgraph*}}\;+\;
\parbox{22mm}{
\begin{fmfgraph*}(22,8)
\fmfset{wiggly_len}{1.5mm}\fmfset{dash_len}{1.5mm}
    \fmfipair{a,b,c,d}
    \fmfiequ{a}{.5[nw,sw]}
    \fmfiequ{b}{.5[ne,se]}
    \fmfiequ{c}{4/11[a,b]}
    \fmfiequ{d}{7/11[a,b]}
    \fmfi{plain}{c--d}
    \fmfi{dashes}{a..c..a}
    \fmfi{dashes}{b..d..b}
\end{fmfgraph*}}\;+\;
\parbox{22mm}{
\begin{fmfgraph*}(22,8)
\fmfset{wiggly_len}{1.5mm}\fmfset{dash_len}{1.5mm}
    \fmfipair{a,b,c,d,e,f,r,s}
    \fmfiequ{a}{.5[nw,sw]}
    \fmfiequ{b}{.5[ne,se]}
    \fmfiequ{c}{4/11[a,b]}
    \fmfiequ{d}{7/11[a,b]}
    \fmfiequ{xpart(e)}{7mm}
    \fmfiequ{xpart(f)}{7mm}
    \fmfiequ{xpart(r)}{15mm}
    \fmfiequ{xpart(s)}{15mm}
    \fmfiequ{ypart(e)}{6.5mm}
    \fmfiequ{ypart(f)}{1.5mm}
    \fmfiequ{ypart(r)}{6.5mm}
    \fmfiequ{ypart(s)}{1.5mm}
    \fmfi{plain}{a..c..a}
    \fmfi{plain}{b..d..b}
    \fmfi{plain}{c--d}
    \fmfiv{d.sh=circle,d.siz=1.5thick}{e}
    \fmfiv{d.sh=circle,d.siz=1.5thick}{f}
    \fmfiv{d.sh=circle,d.siz=1.5thick}{r}
    \fmfiv{d.sh=circle,d.siz=1.5thick}{s}
\end{fmfgraph*}}\;=\\
&=\int_{-1}^0\int_{-1}^0\de\tau
\de\sigma(\puntods{\Delta}+\Delta_{gh})\rvert_\tau\Delta(\puntods{\Delta}+\Delta_{gh})\rvert_\sigma=-\frac{1}{12}\; ,\\{}\\
\mathbf{I}_5&=\;
\parbox{8mm}{
\begin{fmfgraph*}(8,8)
\fmfset{wiggly_len}{1.5mm}\fmfset{dash_len}{1.5mm}
    \fmfipair{a,b,c,d,e,f}
    \fmfiequ{a}{.5[nw,sw]}
    \fmfiequ{b}{.5[ne,se]}
    \fmfiequ{c}{1/4[a,b]}
    \fmfiequ{d}{3/4[a,b]}
    \fmfiequ{xpart(e)}{1mm}
    \fmfiequ{xpart(f)}{7mm}
    \fmfiequ{ypart(e)}{6.5mm}
    \fmfiequ{ypart(f)}{6.5mm}
    \fmfi{plain}{a..b..a}
    \fmfi{plain}{a--b}
    \fmfiv{d.sh=circle,d.siz=1.5thick}{c}
    \fmfiv{d.sh=circle,d.siz=1.5thick}{d}
    \fmfiv{d.sh=circle,d.siz=1.5thick}{e}
    \fmfiv{d.sh=circle,d.siz=1.5thick}{f}
\end{fmfgraph*}}\;-\;
\parbox{8mm}{
\begin{fmfgraph*}(8,8)
\fmfset{wiggly_len}{1.5mm}\fmfset{dash_len}{1.5mm}
    \fmfipair{a,b,c,d}
    \fmfiequ{a}{.5[nw,sw]}
    \fmfiequ{b}{.5[ne,se]}
    \fmfiequ{c}{.5[nw,ne]}
    \fmfiequ{d}{.5[sw,se]}
    \fmfi{plain}{a..d..b}
    \fmfi{dashes}{a..c..b}
    \fmfi{dashes}{a--b}
\end{fmfgraph*}}\;=\int_{-1}^0\int_{-1}^0\de\tau
\de\sigma\Delta(\puntods{\Delta}{}^2-\Delta_{gh}{}^2)=\frac{1}{4}\; ,\\{}\\
\mathbf{I}_6&=\;
\parbox{22mm}{
\begin{fmfgraph*}(22,8)
\fmfset{wiggly_len}{1.5mm}\fmfset{dash_len}{1.5mm}
    \fmfipair{a,b,c,d,e,f,r,s}
    \fmfiequ{a}{.5[nw,sw]}
    \fmfiequ{b}{.5[ne,se]}
    \fmfiequ{c}{4/11[a,b]}
    \fmfiequ{d}{7/11[a,b]}
    \fmfiequ{e}{1/3[c,d]}
    \fmfiequ{xpart(f)}{7mm}
    \fmfiequ{xpart(r)}{15mm}
    \fmfiequ{xpart(s)}{15mm}
    \fmfiequ{ypart(f)}{6.5mm}
    \fmfiequ{ypart(r)}{6.5mm}
    \fmfiequ{ypart(s)}{1.5mm}
    \fmfi{plain}{a..c..a}
    \fmfi{plain}{b..d..b}
    \fmfi{plain}{c--d}
    \fmfiv{d.sh=circle,d.siz=1.5thick}{e}
    \fmfiv{d.sh=circle,d.siz=1.5thick}{f}
    \fmfiv{d.sh=circle,d.siz=1.5thick}{r}
    \fmfiv{d.sh=circle,d.siz=1.5thick}{s}
\end{fmfgraph*}}\;+\;
\parbox{22mm}{
\begin{fmfgraph*}(22,8)
\fmfset{wiggly_len}{1.5mm}\fmfset{dash_len}{1.5mm}
    \fmfipair{a,b,c,d,e,f}
    \fmfiequ{a}{.5[nw,sw]}
    \fmfiequ{b}{.5[ne,se]}
    \fmfiequ{c}{4/11[a,b]}
    \fmfiequ{d}{7/11[a,b]}
    \fmfiequ{e}{1/3[c,d]}
    \fmfiequ{xpart(f)}{7mm}
    \fmfiequ{ypart(f)}{6.5mm}
    \fmfi{plain}{a..c..a}
    \fmfi{plain}{c--d}
    \fmfi{dashes}{b..d..b}
    \fmfiv{d.sh=circle,d.siz=1.5thick}{e}
    \fmfiv{d.sh=circle,d.siz=1.5thick}{f}
\end{fmfgraph*}}\;=\int_{-1}^0\int_{-1}^0\de\tau
\de\sigma\puntos{\Delta}\rvert_\tau\,\puntos{\Delta}(\puntods{\Delta}+\Delta_{gh})\rvert_\sigma=\frac{1}{12}\; , \\{}\\
\mathbf{I}_7&=\;
\parbox{8mm}{
\begin{fmfgraph*}(8,8)
\fmfset{wiggly_len}{1.5mm}\fmfset{dash_len}{1.5mm}
    \fmfipair{a,b,c,d,r,s}
    \fmfiequ{a}{.5[nw,sw]}
    \fmfiequ{b}{.5[ne,se]}
    \fmfiequ{c}{1/4[a,b]}
    \fmfiequ{d}{3/4[a,b]}
    \fmfiequ{xpart(r)}{1mm}
    \fmfiequ{xpart(s)}{7mm}
    \fmfiequ{ypart(r)}{6.5mm}
    \fmfiequ{ypart(s)}{1.5mm}
    \fmfi{plain}{a..b..a}
    \fmfi{plain}{a--b}
    \fmfiv{d.sh=circle,d.siz=1.5thick}{c}
    \fmfiv{d.sh=circle,d.siz=1.5thick}{d}
    \fmfiv{d.sh=circle,d.siz=1.5thick}{r}
    \fmfiv{d.sh=circle,d.siz=1.5thick}{s}
    \end{fmfgraph*}}\;=\int_{-1}^0\int_{-1}^0\de\tau
    \de\sigma\puntos{\Delta}\puntods{\Delta}\puntod{\Delta}=-\frac{1}{12}\; ,\\{}\\
\mathbf{I}_8&=\;
\parbox{22mm}{
\begin{fmfgraph*}(22,8)
\fmfset{wiggly_len}{1.5mm}\fmfset{dash_len}{1.5mm}
    \fmfipair{a,b,c,d,e,f,r,s}
    \fmfiequ{a}{.5[nw,sw]}
    \fmfiequ{b}{.5[ne,se]}
    \fmfiequ{c}{4/11[a,b]}
    \fmfiequ{d}{7/11[a,b]}
    \fmfiequ{e}{1/4[c,d]}
    \fmfiequ{f}{3/4[c,d]}
    \fmfiequ{xpart(r)}{7mm}
    \fmfiequ{xpart(s)}{15mm}
    \fmfiequ{ypart(r)}{6.5mm}
    \fmfiequ{ypart(s)}{6.5mm}
    \fmfi{plain}{a..c..a}
    \fmfi{plain}{b..d..b}
    \fmfi{plain}{c--d}
    \fmfiv{d.sh=circle,d.siz=1.5thick}{e}
    \fmfiv{d.sh=circle,d.siz=1.5thick}{f}
    \fmfiv{d.sh=circle,d.siz=1.5thick}{r}
    \fmfiv{d.sh=circle,d.siz=1.5thick}{s}
\end{fmfgraph*}}\;=\int_{-1}^0\int_{-1}^0\de\tau
\de\sigma\puntos{\Delta}\rvert_\tau\,\puntods{\Delta}\puntod{\Delta}\rvert_\sigma=-\frac{1}{12}\; ,
\end{align*}
\begin{align*}
\mathbf{I}_9&=\;
\parbox{22mm}{
\begin{fmfgraph*}(22,8)
\fmfset{wiggly_len}{1.5mm}\fmfset{dash_len}{1.5mm}
    \fmfipair{a,b,c,d,e,r,s}
    \fmfiequ{a}{.5[nw,sw]}
    \fmfiequ{b}{.5[ne,se]}
    \fmfiequ{c}{4/11[a,b]}
    \fmfiequ{d}{7/11[a,b]}
    \fmfiequ{e}{2/3[c,d]}
    \fmfiequ{xpart(r)}{7mm}
    \fmfiequ{xpart(s)}{7mm}
    \fmfiequ{ypart(r)}{6.5mm}
    \fmfiequ{ypart(s)}{1.5mm}
    \fmfi{plain}{a..c..a}
    \fmfi{plain}{c--d}
    \fmfi{wiggly}{b..d..b}
    \fmfiv{d.sh=circle,d.siz=1.5thick}{e}
    \fmfiv{d.sh=circle,d.siz=1.5thick}{r}
    \fmfiv{d.sh=circle,d.siz=1.5thick}{s}
\end{fmfgraph*}}\;\,+\
\parbox{22mm}{
\begin{fmfgraph*}(22,8)
\fmfset{wiggly_len}{1.5mm}\fmfset{dash_len}{1.5mm}
\fmfipair{a,b,c,d,e}
    \fmfiequ{a}{.5[nw,sw]}
    \fmfiequ{b}{.5[ne,se]}
    \fmfiequ{c}{4/11[a,b]}
    \fmfiequ{d}{7/11[a,b]}
    \fmfiequ{e}{2/3[c,d]}
    \fmfi{dashes}{a..c..a}
    \fmfi{plain}{c--d}
    \fmfi{wiggly}{b..d..b}
    \fmfiv{d.sh=circle,d.siz=1.5thick}{e}
\end{fmfgraph*}}\;\,=\int_{-1}^0\int_{-1}^0\de\tau
\de\sigma(\puntods{\Delta}+\Delta_{gh})\rvert_\tau\,\puntod{\Delta}\Delta_{AF}\rvert_\sigma=0\; , \\{}\\
\mathbf{I}_{10}&=\;
\parbox{22mm}{
\begin{fmfgraph*}(22,8)
\fmfset{wiggly_len}{1.5mm}\fmfset{dash_len}{1.5mm}
 \fmfipair{a,b,c,d,e,f,r}
    \fmfiequ{a}{.5[nw,sw]}
    \fmfiequ{b}{.5[ne,se]}
    \fmfiequ{c}{4/11[a,b]}
    \fmfiequ{d}{7/11[a,b]}
    \fmfiequ{e}{1/4[c,d]}
    \fmfiequ{f}{3/4[c,d]}
    \fmfiequ{xpart(r)}{7mm}
    \fmfiequ{ypart(r)}{6.5mm}
    \fmfi{plain}{a..c..a}
    \fmfi{plain}{c--d}
    \fmfi{wiggly}{b..d..b}
    \fmfiv{d.sh=circle,d.siz=1.5thick}{e}
    \fmfiv{d.sh=circle,d.siz=1.5thick}{f}
    \fmfiv{d.sh=circle,d.siz=1.5thick}{r}
\end{fmfgraph*}}\;\,=\int_{-1}^0\int_{-1}^0\de\tau
\de\sigma\puntos{\Delta}\rvert_\tau\,\puntods{\Delta}\Delta_{AF}\rvert_\sigma=0\; ,\\{}\\
\mathbf{I}_{11}&=\;\,
\parbox{22mm}{
\begin{fmfgraph*}(22,8)
\fmfset{wiggly_len}{1.5mm}\fmfset{dash_len}{1.5mm}
    \fmfipair{a,b,c,d,e,f}
    \fmfiequ{a}{.5[nw,sw]}
    \fmfiequ{b}{.5[ne,se]}
    \fmfiequ{c}{4/11[a,b]}
    \fmfiequ{d}{7/11[a,b]}
    \fmfiequ{e}{1/4[c,d]}
    \fmfiequ{f}{3/4[c,d]}
    \fmfi{wiggly}{a..c..a}
    \fmfi{plain}{c--d}
    \fmfi{wiggly}{b..d..b}
    \fmfiv{d.sh=circle,d.siz=1.5thick}{e}
    \fmfiv{d.sh=circle,d.siz=1.5thick}{f}
\end{fmfgraph*}}\;\,=\int_{-1}^0\int_{-1}^0\de\tau
\de\sigma\Delta_{AF}\rvert_\tau\,\puntods{\Delta}\Delta_{AF}\rvert_\sigma=0\; ,\\{}\\ \mathbf{I}_{12}&=\;\,
\parbox{8mm}{
\begin{fmfgraph*}(8,8)
\fmfset{wiggly_len}{1.5mm}\fmfset{dash_len}{1.5mm}
    \fmfipair{a,b,c,d}
    \fmfiequ{a}{.5[nw,sw]}
    \fmfiequ{b}{.5[ne,se]}
    \fmfiequ{c}{1/4[a,b]}
    \fmfiequ{d}{3/4[a,b]}
    \fmfi{plain}{a--b}
    \fmfi{wiggly}{a..b..a}
    \fmfiv{d.sh=circle,d.siz=1.5thick}{c}
    \fmfiv{d.sh=circle,d.siz=1.5thick}{d}
\end{fmfgraph*}}\;\,=\int_{-1}^0\int_{-1}^0\de\tau
\de\sigma\puntods{\Delta}\Delta_{AF}^2=\mathcal{I}\; ;
\end{align*}
where dots stand for derivatives with respect to the corresponding time variable, straight lines are $qq$ propagators, wiggly lines $\psi\psi$ propagators, dashed lines ghosts propagators and at each vertex corresponds a time integral. We have not found a convenient way to compute $\mathbf{I}_{12}=\mathcal{I}$ directly in the continuum limit.

Then we report the result for flat indices but now in the string inspired case: the twelve diagrams have the same expression as before provided the substitution of any $\Delta$ with $\Delta_{SI}$ and every $\Delta_{gh}$ with $\Delta_{GH}$; the results then are $\mathbf{I_1}=-1/12$, $\mathbf{I_2}=0$, $\mathbf{I_3}=0$, $\mathbf{I_4}=0$, $\mathbf{I_5}=1/6$, $\mathbf{I_6}=0$, $\mathbf{I_7}=0$, $\mathbf{I_8}=0$, $\mathbf{I_9}=0$, $\mathbf{I_{10}}=0$, $\mathbf{I_{11}}=0$, $\mathbf{I_{12}}=1/6$ . Using SI there is no problem in calculating $\mathbf{I}_{12}$ in the continuum limit.

Finally we write down the additional integrals required in the curved indices case:
\begin{align*}
\mi{13}&=\;
\parbox{16mm}{
\begin{fmfgraph*}(16,8)\fmfset{wiggly_len}{1.5mm}\fmfset{dash_len}{1.5mm}\fmfset{zigzag_len}{1mm}\fmfset{zigzag_width}{.7mm}
    \fmfipair{a,b,c,d}
    \fmfiequ{a}{.5[nw,sw]}\fmfiequ{b}{.5[ne,se]}\fmfiequ{c}{.5[a,b]}
    \fmfiequ{xpart(d)}{9mm}\fmfiequ{ypart(d)}{6.5mm}
    \fmfi{plain}{a..c..a}\fmfi{wiggly}{c..b..c}
    \fmfiv{d.sh=circle,d.siz=1.5thick}{d}
\end{fmfgraph*}}\;\,+\;\,
\parbox{16mm}{
\begin{fmfgraph*}(16,8)\fmfset{wiggly_len}{1.5mm}\fmfset{dash_len}{1.5mm}\fmfset{zigzag_len}{1mm}\fmfset{zigzag_width}{.7mm}
    \fmfipair{a,b,c}
    \fmfiequ{a}{.5[nw,sw]}\fmfiequ{b}{.5[ne,se]}\fmfiequ{c}{.5[a,b]}
    \fmfi{plain}{a..c..a}\fmfi{zigzag}{c..b..c}
\end{fmfgraph*}}\;=\int_{-1}^0\ide{\tau}\Delta\rvert_\tau(\puntos{\Delta}_{AF}+\Delta_{FG})\rvert_\tau=0\; ,\\{}\\
\mi{14}&=\;
\parbox{22mm}{
\begin{fmfgraph*}(22,8)\fmfset{wiggly_len}{1.5mm}\fmfset{dash_len}{1.5mm}\fmfset{zigzag_len}{1mm}\fmfset{zigzag_width}{.7mm}
    \fmfipair{a,b,c,d,e,f,g}
    \fmfiequ{a}{.5[nw,sw]}\fmfiequ{d}{.5[ne,se]}\fmfiequ{b}{4/11[a,d]}\fmfiequ{c}{7/11[a,d]}
    \fmfiequ{e}{1/4[b,c]}
    \fmfiequ{xpart(f)}{7mm}\fmfiequ{ypart(f)}{6.5mm}
    \fmfiequ{xpart(g)}{15mm}\fmfiequ{ypart(g)}{6.5mm}
    \fmfi{plain}{a..b..a}\fmfi{wiggly}{c..d..c}\fmfi{plain}{b--c}
    \fmfiv{d.sh=circle,d.siz=1.5thick}{e}\fmfiv{d.sh=circle,d.siz=1.5thick}{f}\fmfiv{d.sh=circle,d.siz=1.5thick}{g}
\end{fmfgraph*}}\;\,+\;\,
\parbox{22mm}{
\begin{fmfgraph*}(22,8)\fmfset{wiggly_len}{1.5mm}\fmfset{dash_len}{1.5mm}\fmfset{zigzag_len}{1mm}\fmfset{zigzag_width}{.7mm}
    \fmfipair{a,b,c,d,e,f}
    \fmfiequ{a}{.5[nw,sw]}\fmfiequ{d}{.5[ne,se]}\fmfiequ{b}{4/11[a,d]}\fmfiequ{c}{7/11[a,d]}
    \fmfiequ{e}{1/4[b,c]}
    \fmfiequ{xpart(f)}{7mm}\fmfiequ{ypart(f)}{6.5mm}
    \fmfi{plain}{a..b..a}\fmfi{zigzag}{c..d..c}\fmfi{plain}{b--c}
    \fmfiv{d.sh=circle,d.siz=1.5thick}{e}\fmfiv{d.sh=circle,d.siz=1.5thick}{f}
\end{fmfgraph*}}\;\,=\int_{-1}^0\int_{-1}^0\de\tau\ide{\sigma}\puntos{\Delta}\rvert_\tau\:\puntos{\Delta}(\puntos{\Delta}_{AF}+\Delta_{FG})\rvert_\sigma=0\; ,\\{}\\
\mi{15}&=\;
\parbox{22mm}{
\begin{fmfgraph*}(22,8)\fmfset{wiggly_len}{1.5mm}\fmfset{dash_len}{1.5mm}\fmfset{zigzag_len}{1mm}\fmfset{zigzag_width}{.7mm}
    \fmfipair{a,b,c,d,e,f,g}
    \fmfiequ{a}{.5[nw,sw]}\fmfiequ{d}{.5[ne,se]}\fmfiequ{b}{4/11[a,d]}\fmfiequ{c}{7/11[a,d]}
    \fmfiequ{xpart(e)}{7mm}\fmfiequ{ypart(e)}{1.5mm}
    \fmfiequ{xpart(f)}{7mm}\fmfiequ{ypart(f)}{6.5mm}
    \fmfiequ{xpart(g)}{15mm}\fmfiequ{ypart(g)}{6.5mm}
    \fmfi{plain}{a..b..a}\fmfi{wiggly}{c..d..c}\fmfi{plain}{b--c}
    \fmfiv{d.sh=circle,d.siz=1.5thick}{e}\fmfiv{d.sh=circle,d.siz=1.5thick}{f}\fmfiv{d.sh=circle,d.siz=1.5thick}{g}
\end{fmfgraph*}}\;\,+\;\,
\parbox{22mm}{
\begin{fmfgraph*}(22,8)\fmfset{wiggly_len}{1.5mm}\fmfset{dash_len}{1.5mm}\fmfset{zigzag_len}{1mm}\fmfset{zigzag_width}{.7mm}
    \fmfipair{a,b,c,d,e,f}
    \fmfiequ{a}{.5[nw,sw]}\fmfiequ{d}{.5[ne,se]}\fmfiequ{b}{4/11[a,d]}\fmfiequ{c}{7/11[a,d]}
    \fmfiequ{xpart(e)}{7mm}\fmfiequ{ypart(e)}{1.5mm}
    \fmfiequ{xpart(f)}{7mm}\fmfiequ{ypart(f)}{6.5mm}
    \fmfi{plain}{a..b..a}\fmfi{zigzag}{c..d..c}\fmfi{plain}{b--c}
    \fmfiv{d.sh=circle,d.siz=1.5thick}{e}\fmfiv{d.sh=circle,d.siz=1.5thick}{f}
\end{fmfgraph*}}\;\,=\int_{-1}^0\int_{-1}^0\de\tau\ide{\sigma}\puntods{\Delta}\rvert_\tau\Delta(\puntos{\Delta}_{AF}+\Delta_{FG})\rvert_\sigma=0\; ,\\{}\\
\mi{16}&=\;
\parbox{22mm}{
\begin{fmfgraph*}(22,8)\fmfset{wiggly_len}{1.5mm}\fmfset{dash_len}{1.5mm}\fmfset{zigzag_len}{1mm}\fmfset{zigzag_width}{.7mm}
    \fmfipair{a,b,c,d,f,g}
    \fmfiequ{a}{.5[nw,sw]}\fmfiequ{d}{.5[ne,se]}\fmfiequ{b}{4/11[a,d]}\fmfiequ{c}{7/11[a,d]}
    \fmfiequ{xpart(f)}{7mm}\fmfiequ{ypart(f)}{6.5mm}
    \fmfiequ{xpart(g)}{15mm}\fmfiequ{ypart(g)}{6.5mm}
    \fmfi{wiggly}{a..b..a}\fmfi{wiggly}{c..d..c}\fmfi{plain}{b--c}
    \fmfiv{d.sh=circle,d.siz=1.5thick}{f}\fmfiv{d.sh=circle,d.siz=1.5thick}{g}
\end{fmfgraph*}}\;\,+\,\;
\parbox{22mm}{
\begin{fmfgraph*}(22,8)\fmfset{wiggly_len}{1.5mm}\fmfset{dash_len}{1.5mm}\fmfset{zigzag_len}{1mm}\fmfset{zigzag_width}{.7mm}
    \fmfipair{a,b,c,d}
    \fmfiequ{a}{.5[nw,sw]}\fmfiequ{d}{.5[ne,se]}\fmfiequ{b}{4/11[a,d]}\fmfiequ{c}{7/11[a,d]}
    \fmfi{zigzag}{a..b..a}\fmfi{zigzag}{c..d..c}\fmfi{plain}{b--c}
\end{fmfgraph*}}\;\,+2\;\,
\parbox{22mm}{
\begin{fmfgraph*}(22,8)\fmfset{wiggly_len}{1.5mm}\fmfset{dash_len}{1.5mm}\fmfset{zigzag_len}{1mm}\fmfset{zigzag_width}{.7mm}
    \fmfipair{a,b,c,d,f}
    \fmfiequ{a}{.5[nw,sw]}\fmfiequ{d}{.5[ne,se]}\fmfiequ{b}{4/11[a,d]}\fmfiequ{c}{7/11[a,d]}
    \fmfiequ{xpart(f)}{7mm}\fmfiequ{ypart(f)}{6.5mm}
    \fmfi{wiggly}{a..b..a}\fmfi{zigzag}{c..d..c}\fmfi{plain}{b--c}
    \fmfiv{d.sh=circle,d.siz=1.5thick}{f}
\end{fmfgraph*}}\\{}\\
&= \int_{-1}^0\int_{-1}^0 \de\tau
\de\sigma\,(\puntod{\Delta}_{AF}+\Delta_{FG})\rvert_\tau\,\Delta\,(\puntod{\Delta}_{AF}+\Delta_{FG})\rvert_\sigma=0\; ,\\{}\\
\mi{17}&=\;\,
\parbox{22mm}{
\begin{fmfgraph*}(22,8)\fmfset{wiggly_len}{1.5mm}\fmfset{dash_len}{1.5mm}\fmfset{zigzag_len}{1mm}\fmfset{zigzag_width}{.7mm}
    \fmfipair{a,b,c,d,e}
    \fmfiequ{a}{.5[nw,sw]}\fmfiequ{d}{.5[ne,se]}\fmfiequ{b}{4/11[a,d]}\fmfiequ{c}{7/11[a,d]}
    \fmfiequ{xpart(e)}{7mm}\fmfiequ{ypart(e)}{6.5mm}
    \fmfi{wiggly}{a..b..a}\fmfi{plain}{b--c}\fmfi{dashes}{c..d..c}
    \fmfiv{d.sh=circle,d.siz=1.5thick}{e}
\end{fmfgraph*}}\;\,+\,\;
\parbox{22mm}{
\begin{fmfgraph*}(22,8)\fmfset{wiggly_len}{1.5mm}\fmfset{dash_len}{1.5mm}\fmfset{zigzag_len}{1mm}\fmfset{zigzag_width}{.7mm}
    \fmfipair{a,b,c,d}
    \fmfiequ{a}{.5[nw,sw]}\fmfiequ{d}{.5[ne,se]}\fmfiequ{b}{4/11[a,d]}\fmfiequ{c}{7/11[a,d]}
    \fmfi{zigzag}{a..b..a}\fmfi{plain}{b--c}\fmfi{dashes}{c..d..c}
\end{fmfgraph*}}\;=\int_{-1}^0\int_{-1}^0\de\tau\ide{\sigma}(\puntos{\Delta}_{AF}+\Delta_{FG})\rvert_\tau\Delta\Delta_{gh}\rvert_\sigma=0\; ,
\end{align*}
\begin{align*}
\mi{18}&=\;
\parbox{22mm}{
\begin{fmfgraph*}(22,8)\fmfset{wiggly_len}{1.5mm}\fmfset{dash_len}{1.5mm}\fmfset{zigzag_len}{1mm}\fmfset{zigzag_width}{.7mm}
    \fmfipair{a,b,c,d,e,f}
    \fmfiequ{a}{.5[nw,sw]}\fmfiequ{d}{.5[ne,se]}\fmfiequ{b}{4/11[a,d]}\fmfiequ{c}{7/11[a,d]}
    \fmfiequ{e}{3/4[b,c]}
    \fmfiequ{xpart(f)}{7mm}\fmfiequ{ypart(f)}{6.5mm}
    \fmfi{wiggly}{a..b..a}\fmfi{plain}{b--c}\fmfi{wiggly}{c..d..c}
    \fmfiv{d.sh=circle,d.siz=1.5thick}{e}\fmfiv{d.sh=circle,d.siz=1.5thick}{f}
\end{fmfgraph*}}\;\,+\;\,
\parbox{22mm}{
\begin{fmfgraph*}(22,8)\fmfset{wiggly_len}{1.5mm}\fmfset{dash_len}{1.5mm}\fmfset{zigzag_len}{1mm}\fmfset{zigzag_width}{.7mm}
    \fmfipair{a,b,c,d,e}
    \fmfiequ{a}{.5[nw,sw]}\fmfiequ{d}{.5[ne,se]}\fmfiequ{b}{4/11[a,d]}\fmfiequ{c}{7/11[a,d]}
    \fmfiequ{e}{3/4[b,c]}
    \fmfi{zigzag}{a..b..a}\fmfi{plain}{b--c}\fmfi{wiggly}{c..d..c}
    \fmfiv{d.sh=circle,d.siz=1.5thick}{e}
\end{fmfgraph*}}\;\,=\int_{-1}^0\int_{-1}^0\de\tau\ide{\sigma}(\puntod{\Delta}_{AF}+\Delta_{FG})\rvert_\tau\puntod{\Delta}\Delta_{AF}\rvert_\sigma=0\; ,\\{}\\
\mi{19}&=\;\parbox{8mm}{
\begin{fmfgraph*}(8,8)
\fmfset{wiggly_len}{1.5mm}\fmfset{dash_len}{1.5mm}
    \fmfipair{a,b,c,d}
    \fmfiequ{a}{.5[nw,sw]}
    \fmfiequ{b}{.5[ne,se]}
    \fmfiequ{xpart(c)}{1mm}
    \fmfiequ{ypart(c)}{6.5mm}
    \fmfiequ{xpart(d)}{7mm}
    \fmfiequ{ypart(d)}{6.5mm}
    \fmfi{plain}{a--b}
    \fmfi{wiggly}{a..b..a}
    \fmfiv{d.sh=circle,d.siz=1.5thick}{c}
    \fmfiv{d.sh=circle,d.siz=1.5thick}{d}
\end{fmfgraph*}}\;-\;\parbox{8mm}{
\begin{fmfgraph*}(8,8)
\fmfset{wiggly_len}{1.5mm}\fmfset{dash_len}{1.5mm}
    \fmfipair{a,b,c,d}
    \fmfiequ{a}{.5[nw,sw]}
    \fmfiequ{b}{.5[ne,se]}
    \fmfiequ{xpart(c)}{1mm}
    \fmfiequ{ypart(c)}{1.5mm}
    \fmfiequ{xpart(d)}{7mm}
    \fmfiequ{ypart(d)}{6.5mm}
    \fmfi{plain}{a--b}
    \fmfi{wiggly}{a..b..a}
    \fmfiv{d.sh=circle,d.siz=1.5thick}{c}
    \fmfiv{d.sh=circle,d.siz=1.5thick}{d}
\end{fmfgraph*}}\; - 2\;\:
\parbox{8mm}{
\begin{fmfgraph*}(8,8)
\fmfset{wiggly_len}{1.5mm}\fmfset{dash_len}{1.5mm}\fmfset{zigzag_len}{1mm}\fmfset{zigzag_width}{.7mm}
    \fmfipair{a,b}
    \fmfiequ{a}{.5[nw,sw]}
    \fmfiequ{b}{.5[ne,se]}
    \fmfi{plain}{a--b}
    \fmfi{zigzag}{a..b..a}
\end{fmfgraph*}}=\int_{-1}^0\int_{-1}^0\de\tau\ide{\sigma}\Delta(\puntods{\Delta}_{AF}\Delta_{AF}-{\puntos{\Delta}_{AF}} {\puntod{\Delta}}_{AF}-2\Delta_{FG}^2) = \frac{1}{12}\; ;
\end{align*}
where the zig-zag lines stand for $\alpha$ propagators.

To perform all these integrals we used the following relations, valid in mode regularization, \emph{i.e.} for finite $M$:
\begin{gather*}
  \Delta_{gh}=\puntoss{\Delta}\quad,\\
  \left(\puntods{\Delta}+\puntoss{\Delta}\right)\rvert_\tau =\partial_\tau\left(\puntos{\Delta}\rvert_\tau\right)\quad,
\puntos{\Delta}_{SI}\rvert_\tau =0\quad,\\
\puntods{\Delta}_{SI} + \Delta_{GH} = 1\quad,
\Delta_{AF}\rvert_{\tau}=0\quad,\\
\puntod{\Delta}_{AF} + \Delta_{FG} = 0\quad,\\
\puntoss{\Delta}_{SI} = e^{-i\pi x}\puntod{\Delta}_{AF} - e^{-2\pi i (M+1)x}-1 \quad .
\end{gather*}
 For a detailed discussion of the techniques used in solving such integrals see \cite{Bastianelli:book}.
\acknowledgments
We would like to thank Fiorenzo Bastianelli and Olindo Corradini that provided us with helpful suggestions.
\bibliography{abf}

\begin{thebibliography}{10}

\bibitem{Berezin:1975eg}
F.~A. Berezin and M.~S. Marinov,
\newblock JETP Lett. {\bf 21}, 320 (1975).

\bibitem{Brink:1976sz}
L.~Brink, S.~Deser, B.~Zumino, P.~Di~Vecchia, and P.~S. Howe,
\newblock Phys. Lett. {\bf B64}, 435 (1976).

\bibitem{Barducci:1976qu}
A.~Barducci, R.~Casalbuoni, and L.~Lusanna,
\newblock Nuovo Cim. {\bf A35}, 377 (1976).

\bibitem{Gershun:1979fb}
V.~D. Gershun and V.~I. Tkach,
\newblock JETP Lett. {\bf 29}, 288 (1979).

\bibitem{Howe:1989vn}
P.~S. Howe, S.~Penati, M.~Pernici, and P.~K. Townsend,
\newblock Class. Quant. Grav. {\bf 6}, 1125 (1989).

\bibitem{AlvarezGaume:1983at}
L.~Alvarez-Gaume,
\newblock Commun. Math. Phys. {\bf 90}, 161 (1983).

\bibitem{AlvarezGaume:1983ig}
L.~Alvarez-Gaume and E.~Witten,
\newblock Nucl. Phys. {\bf B234}, 269 (1984).

\bibitem{Friedan:1983xr}
D.~Friedan and P.~Windey,
\newblock Nucl. Phys. {\bf B235}, 395 (1984).

\bibitem{Bastianelli:1991be}
F.~Bastianelli,
\newblock Nucl. Phys. {\bf B376}, 113 (1992), hep-th/9112035.

\bibitem{Bastianelli:1992ct}
F.~Bastianelli and P.~van Nieuwenhuizen,
\newblock Nucl. Phys. {\bf B389}, 53 (1993), hep-th/9208059.

\bibitem{Bastianelli:book}
F.~Bastianelli and P.~van Nieuwenhuizen,
\newblock {\em Path Integrals and Anomalies in Curved Space} (Cambridge
  University Press, 2006).

\bibitem{Strassler:1992zr}
M.~J. Strassler,
\newblock Nucl. Phys. {\bf B385}, 145 (1992), hep-ph/9205205.

\bibitem{Schmidt:1993rk}
M.~G. Schmidt and C.~Schubert,
\newblock Phys. Lett. {\bf B318}, 438 (1993), hep-th/9309055.

\bibitem{Schmidt:1994zj}
M.~G. Schmidt and C.~Schubert,
\newblock Phys. Lett. {\bf B331}, 69 (1994), hep-th/9403158.

\bibitem{D'Hoker:1995ax}
E.~D'Hoker and D.~G. Gagne,
\newblock Nucl. Phys. {\bf B467}, 272 (1996), hep-th/9508131.

\bibitem{D'Hoker:1995bj}
E.~D'Hoker and D.~G. Gagne,
\newblock Nucl. Phys. {\bf B467}, 297 (1996), hep-th/9512080.

\bibitem{Bastianelli:2002fv}
F.~Bastianelli and A.~Zirotti,
\newblock Nucl. Phys. {\bf B642}, 372 (2002), hep-th/0205182.

\bibitem{Bastianelli:2002qw}
F.~Bastianelli, O.~Corradini, and A.~Zirotti,
\newblock Phys. Rev. {\bf D67}, 104009 (2003), hep-th/0211134.

\bibitem{Bastianelli:2005vk}
F.~Bastianelli, P.~Benincasa, and S.~Giombi,
\newblock JHEP {\bf 04}, 010 (2005), hep-th/0503155.

\bibitem{Bastianelli:2005uy}
F.~Bastianelli, P.~Benincasa, and S.~Giombi,
\newblock JHEP {\bf 10}, 114 (2005), hep-th/0510010.

\bibitem{Schubert:2001he}
C.~Schubert,
\newblock Phys. Rept. {\bf 355}, 73 (2001), hep-th/0101036.

\bibitem{Bastianelli:2004zp}
F.~Bastianelli and C.~Schubert,
\newblock JHEP {\bf 02}, 069 (2005), gr-qc/0412095.

\bibitem{Bastianelli:2007jv}
F.~Bastianelli, U.~Nucamendi, C.~Schubert, and V.~M. Villanueva,
\newblock JHEP {\bf 11}, 099 (2007), 0710.5572.

\bibitem{Hollowood:2007kt}
T.~J. Hollowood and G.~M. Shore,
\newblock Phys. Lett. {\bf B655}, 67 (2007), 0707.2302.

\bibitem{Hollowood:2007ku}
T.~J. Hollowood and G.~M. Shore,
\newblock Nucl. Phys. {\bf B795}, 138 (2008), 0707.2303.

\bibitem{Bastianelli:1998jm}
F.~Bastianelli, K.~Schalm, and P.~van Nieuwenhuizen,
\newblock Phys. Rev. {\bf D58}, 044002 (1998), hep-th/9801105.

\bibitem{Berezin:1971jf}
F.~A. Berezin,
\newblock Teor. Mat. Fiz. {\bf 6}, 194 (1971).

\bibitem{Mizrahi:1975pw}
M.~M. Mizrahi,
\newblock J. Math. Phys. {\bf 16}, 2201 (1975).

\bibitem{Gervais:1976ws}
J.-L. Gervais and A.~Jevicki,
\newblock Nucl. Phys. {\bf B110}, 93 (1976).

\bibitem{deBoer:1995hv}
J.~De~Boer, B.~Peeters, K.~Skenderis, and P.~Van~Nieuwenhuizen,
\newblock Nucl. Phys. {\bf B446}, 211 (1995), hep-th/9504097.

\bibitem{DeBoer:1995cb}
J.~de~Boer, B.~Peeters, K.~Skenderis, and P.~van Nieuwenhuizen,
\newblock Nucl. Phys. {\bf B459}, 631 (1996), hep-th/9509158.

\bibitem{Kleinert:1999aq}
H.~Kleinert and A.~Chervyakov,
\newblock Phys. Lett. {\bf B464}, 257 (1999), hep-th/9906156.

\bibitem{Bastianelli:2000pt}
F.~Bastianelli, O.~Corradini, and P.~van Nieuwenhuizen,
\newblock Phys. Lett. {\bf B490}, 154 (2000), hep-th/0007105.

\bibitem{Bastianelli:2000nm}
F.~Bastianelli, O.~Corradini, and P.~van Nieuwenhuizen,
\newblock Phys. Lett. {\bf B494}, 161 (2000), hep-th/0008045.

\bibitem{Schalm:1998ix}
K.~Schalm and P.~van Nieuwenhuizen,
\newblock Phys. Lett. {\bf B446}, 247 (1999), hep-th/9810115.

\bibitem{Bastianelli:2003bg}
F.~Bastianelli, O.~Corradini, and A.~Zirotti,
\newblock JHEP {\bf 01}, 023 (2004), hep-th/0312064.

\end{thebibliography}
\end{fmffile}
\end{document}